\documentclass[12pt,a4paper]{article}
\usepackage{amsmath}
\usepackage{graphicx}
\usepackage{amssymb}
\usepackage[normalem]{ulem}
\usepackage{cancel}
\usepackage{color}
\usepackage{xspace}
\newcommand{\BDB}{two-process\xspace}

\newcommand{\Qmax}{Q_{\rm max}}
\newcommand{\Qs}{Q_{S}}
\begin{document}
\title{Mathematical Models for Sleep-Wake Dynamics:  Comparison of the Two-Process
Model and a Mutual Inhibition Neuronal Model}
\author{A.C.~Skeldon$^*$, D.-J.~Dijk$^\dag$, G.~Derks$^*$ \\
$^*$ \small Department of Mathematics, University of Surrey, Guildford, Surrey, GU2 7XH, \\
$^\dag$ \small Faculty of Health and Medical Sciences, University of Surrey, Guildford, Surrey, GU2 7XH. 
}
\maketitle
\begin{abstract}
Sleep is essential for the maintenance of the brain and the body, yet many features of sleep are 
poorly understood and mathematical models
are an important tool for probing proposed biological mechanisms.
The most well-known mathematical model of sleep regulation,
the two-process model, models the sleep-wake cycle by two oscillators: a circadian oscillator 
and a homeostatic oscillator.
An alternative, more recent, model considers the
mutual inhibition
of sleep promoting neurons and the ascending arousal system regulated by 
homeostatic and circadian processes.

Here we show there are fundamental similarities between these two
models. The implications are illustrated with two important sleep-wake
phenomena.  Firstly, we show that in the two-process model,
transitions between different numbers of daily sleep episodes can be
classified as grazing bifurcations.  This provides the theoretical
underpinning for numerical results showing that the sleep patterns of
many mammals can be explained by the 
    mutual inhibition model.  Secondly, we show that when
sleep deprivation disrupts the sleep-wake cycle, ostensibly different
measures of sleepiness in the two models are closely related.

The demonstration of the mathematical similarities of the two models
is valuable because not only does it allow some features of the two-process model
to be interpreted physiologically but it also
means that knowledge
gained from study of the two-process
model can be used to inform understanding of the behaviour of the mutual
inhibition model.  This is important because the mutual inhibition model
and its extensions are
increasingly being used as a tool to understand a diverse range of sleep-wake
phenomena such as the design of optimal shift-patterns, yet the values
it uses for parameters associated with the circadian and homeostatic processes
are very different
from those that have been experimentally measured in the context of the
two-process model.

\end{abstract}
\section{Background}

Reduced or mis-timed sleep is increasingly recognized as presenting
a significant health risk and has been correlated with increases in
a diverse range of medical problems including all-cause mortality,
cardio-vascular disease, diabetes and
impaired vigilance and cognition
\cite{Cappuccioetal_2010,Knutson_2010,Cappuccioetal_2011,Nielsenetal_2011,
Kronholm_11}.  
The biological mechanisms that result in such problems are
beginning to be understood:  recent work 
has shown that changes to the duration or timing of the human
sleep-wake cycle can result in the up- or down- regulation and changes to the
temporal pattern
of large numbers of genes, significantly altering the behaviour of
genes associated with biological processes including metabolic,
inflammatory, immune and stress responses  and circadian rhythmicity
\cite{MollerLevetetal_2013,Archeretal_2014}.
To further understand the underlying phenomena and associations that 
govern sleep-wake regulation,
mathematical models are an important tool to help
clarify concepts, challenge accepted ideas and aid in the interpretation
of data.

A review of early mathematical models of sleep is given in 
\cite{Czeisler_84}, leading up to the seminal model of
Borb{\'e}ly, Daan and Beersma 
\cite{Borbely_82, Daan_84},
usually called the \BDB model, and extended by Borb{\'e}ly and Achermann  \cite{BorbelyAchermann_99}. 
As indicated by its name, the \BDB model
proposes that the sleep-wake cycle can be
understood in terms of 
two processes, a homeostatic process and a circadian process.  The homeostatic
process takes the form of a relaxation oscillator that results in a
monotonically increasing `sleep pressure' during wake that is dissipated
during sleep. Switching from wake to sleep and from sleep to wake 
occurs at upper and lower
threshold values of the sleep pressure respectively, where the thresholds 
are modulated by an approximately 
sinusoidal circadian oscillator.  This model has proved compelling for
both its physiological grounding and its graphical simplicity and has 
been used extensively (there are over 1500 citations to \cite{Borbely_82}
and 600 citations to \cite{Daan_84}
to-date). For example: to explain 
why only a relatively short period of recovery 
sleep is needed to compensate for even lengthy periods of sleep deprivation
\cite{Borbely_82}; to explain chronotype changes in adolescents \cite{HL_HB_13}. 
Extensions of the \BDB model have been developed
to explain the results of chronic sleep
restriction experiments \cite{Avinash_05,McCauley_09}. 
 Despite its success, it remains 
difficult to relate the threshold values in the \BDB model and its extensions
to physiological processes.

Advances in neurophysiology have led to a proliferation
of models that aim to extend the \BDB model to a more physiological
setting
\cite{Tamakawa_etal_2006,Behn_07,PR_JBR_07,PVB_JBR_2009,Behn_10,RBT_10,Booth_12}.
A recent review is given in \cite{Booth_14}.
The most extensively tested of these is the model of Phillips and
Robinson \cite{PR_JBR_07} (the PR model), which has been used to
explain sleep fragmentation experiments \cite{FPR_PRE_08}, differences
in mammalian sleep patterns \cite{PRKA_PLoSCB_10} and subjective
fatigue during sleep deprivation \cite{FPR_JTB_10}.  The PR model has
also been extended to allow for the inclusion of the effects of
caffeine \cite{PFPR_JTB_11} and to allow for feedback of the
sleep-wake cycle on the circadian oscillator in order to explain
spontaneous internal desynchrony \cite{PCR_JBR_10,PCK_JBR_11}.

In~\cite{PR_JBR_07,RBT_10}, it was observed that the results of
  their physiologically based models could be presented in a
  qualitatively similar way as those from the \BDB model.  
Here we show that some features of the PR model are not only
qualitatively, but also quantitatively similar to the \BDB model: the
parameters in the PR model can be explicitly related to the parameters
in the \BDB model, giving a physiological interpretation to the
thresholds in the \BDB model.  We illustrate the consequences
of this explicit relation with two important
sleep-wake phenomena.  First, by using the fact that the
\BDB model can be represented as a one-dimensional map with
discontinuities~\cite{nakao_etal_im_1997,nakao_yamamoto_pcn_1998},
we demonstrate how transitions between monophasic and polyphasic sleep
occur through grazing bifurcations.  These grazing bifurcations
are then used to provide a theoretical underpinning for
observations that many mammalian sleep
patterns can be understood within a common framework by varying just
two parameters in the PR model \cite{PRKA_PLoSCB_10}.  
Second, turning to sleep deprivation experiments, we
show how the `wake effort' concept introduced in the PR model to
explain sleep deprivation can be explicitly related to the \BDB model.
This shows that the wake effort is closely related to the difference
between the homeostatic pressure and the circadian oscillator, a 
measure often used in the context of the two-process model to understand
sleepiness.
Furthermore we discuss briefly how the PR model may explain effects of
chronic partial sleep deprivation on waking performance.

\section{Sleep models}
\subsection{The two process model}
\begin{figure}[htb]
\includegraphics[scale=0.6]{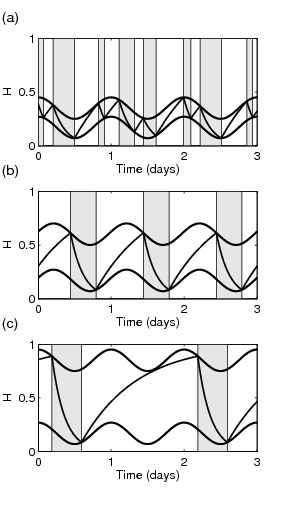}
\caption{Sleep-wake cycles generated by the 
\BDB model. $C(t)=\sin(\omega t), H_0^-=0.17, a=0.10, \chi_s=4.2 \mbox{hrs}, 
\chi_w=18.2 \mbox{hrs},
\mu=1$.  (a) $H_0^+ = 0.35$, (b) $H_0^+ = 0.60$, (c) $H_0^+ = 0.85$.  
Parameters as in \cite{Daan_84}, Figure 3.  
The times when sleep occurs ($H$ decreasing) are shaded. 
}
\label{fig:BA_Daanparam}
\end{figure}

The \BDB model considers a homeostatic pressure $H(t)$ that decreases
exponentially during sleep,
\begin{equation}
H(t) = H_0 \rm{e}^{(t_0-t)/\chi_s} \label{eq:BAsleep}
\end{equation}
and increases during wake,
\begin{equation}
H = \mu + (H_0 - \mu) \rm{e}^{(t_0-t)/\chi_w}.  \label{eq:BAwake}
\end{equation}
The parameter $\mu$ is known as
the `upper asymptote' \cite{Avinash_05,McCauley_09},
 this is the value that the homeostatic pressure $H$ would reach 
if no switch to sleep occurred.  Similarly there is a `lower asymptote'
of zero. 
Switching between wake and sleep occurs when the homeostatic
pressure $H(t)$ reaches an
upper threshold, $H^+(t)$, that consists of a mean value $H^+_0$ 
modulated by a circadian process $C(t)$,
\begin{equation}
H^+(t) = H^+_0 + a C(t).
\label{eq:BAupperthreshold}
\end{equation}
The switch between sleep and wake occurs when
$H(t)$ reaches a lower threshold, $H^-(t)$, 
\begin{equation}
H^-(t) = H^-_0 + a C(t),
\label{eq:BAlowerthreshold}
\end{equation}
where 
$C(t)$ is a periodic function of period 24 hours. In the simplest
cases 
$$C(t)=\sin(\omega (t-\alpha)),$$
but more complicated forms that include higher harmonics, such as  
\begin{eqnarray*}
C(t)  & = &  0.97 \sin \omega (t-\alpha) + 0.22 \sin 2 \omega (t-\alpha) 
+ 0.007 \sin 3 \omega (t-\alpha) \\
& &  + 0.03 \sin 4 \omega (t-\alpha) + 0.001 \sin 5 \omega (t-\alpha),
\end{eqnarray*}
have also been used \cite{BorbelyAchermann_99}.  
Typical results of this model illustrating its rich dynamics 
are shown in Figure~\ref{fig:BA_Daanparam}.  

\newpage

\subsection{Phillips and Robinson model (PR model)}
\begin{figure}[htb]
\includegraphics[scale=1.0]{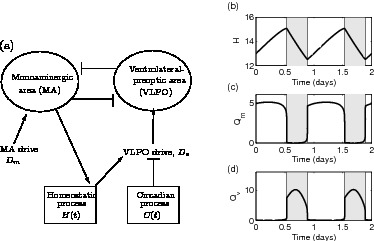}
\caption{(a) Diagrammatic description of the PR model showing
the links between the VLPO, MA, the homeostatic and the circadian
processes. 
(b), (c) and (d) show typical timeseries for the level of the homeostat, $H$,
and the firing rates of the MA and VLPO, $Q_m$ and $Q_v$, respectively.
The times where sleep occurs are shaded. 
}
\label{fig:PRmodel}
\end{figure}

At the core of the PR model are two groups of neurons:
mono-aminergic (MA) neurons in the ascending arousal system that promote wake 
and neurons based in the ventro-lateral 
pre-optic (VLPO) area of the hypothalamus that promote sleep.  Phillips
and Robinson model the 
interaction between the MA and the VLPO as mutually inhibitory. In the
absence of further effects, this would mean that the model would either stay in a state
with the MA active (wake) or in a state with the VLPO active (sleep) and no switching
between the states would occur.    Switching between sleep and wake occurs because the
model also includes a drive to the VLPO that is time dependent and consists of two
components:  a circadian drive, $C(t)$, and a homeostatic drive $H(t)$. 
 The structure of the PR model is shown in Figure~\ref{fig:PRmodel}(a).

The neurons are modelled at a population level and are represented by their
mean cell body potential relative to rest, $V_j$ for $j= m, v$, 
where $v$ represents the VLPO group and $m$ represents the MA. The potential
is related to the firing rates of the neurons by the firing
function, $Q_j$, 
\begin{eqnarray}
Q_j = \frac{\Qmax}{1+\exp[-(V-\theta)/\sigma ']},
\label{eq:softswitch}
\end{eqnarray}
where $\Qmax$ is the maximum firing rate and $\theta$ is the mean firing threshold 
relative to resting. The function $Q_j$ is a sigmoid function, which is close to zero for all 
negative values of $V_j$ and then saturates exponentially fast to~$\Qmax$.

The neuronal dynamics are represented by
\begin{eqnarray}
\tau_v \dot{V}_v+V_v&=&-\nu_{vm}Q_m+D_v, \nonumber \\
\tau_m \dot{V}_m+V_m&=&-\nu_{mv}Q_v+D_m, \label{eq:PRmodel}
\end{eqnarray}
where the drive to the VLPO, $D_v$ and to the MA, $D_m$ are given by
\begin{eqnarray*}
D_v &=& \nu_{vh}H - \nu_{vc} C - A_v,\\
D_m &=& A_m.
\end{eqnarray*}
The homeostatic component of the drive, $H$ is modelled by
\begin{eqnarray}\label{Hprime}
\chi \dot{H} + H = \bar \mu Q_m,
\end{eqnarray}
and the circadian drive, $C$, is approximated by 
\begin{eqnarray*}
C(t)=\cos(\omega (t-\alpha)),
\end{eqnarray*}
where  $\omega=2\pi/24$ hrs$^{-1}$
and $\alpha$ is a shift that specifies the distance from the 
circadian maximum.  
Typically, $\alpha$ is chosen so that the
switch from sleep to wake occurs at an appropriate clock time.

Typical results produced by the PR model are shown in Figure~\ref{fig:PRmodel}(b)-(d).
During wake, the firing rate of the MA neurons is high ($Q_m\approx
  5$ s$^{-1}$), that of the 
VLPO is low and the homeostatic pressure tends to increase, while during
sleep the firing rate of the MA neurons is low ($Q_m\approx0$ s$^{-1}$), 
that of the VLPO is high
and the homeostatic pressure tends to decrease.  
Note that in the PR model
switching between wake and sleep is defined to occur when $Q_m$ reaches
the threshold value of one;  this 
differs from the timing of the 
maximum and minimum homeostatic pressure by a few
minutes. Obviously, the exact choice of the threshold does not
  play an important role in the dynamics of the system, but does change
  the regions that are labelled as sleep or wake.

\newpage
\section{Comparison of the PR and \BDB models}\label{sect:BAPRequivalence}

As recognised in \cite{FPR_PRE_08}, 
since changes in neuronal potentials happen much faster
than changes associated with the homeostatic pressure,
$\tau_j \ll \chi$,  there is a strong separation of timescales
in the PR model. 
This strong separation of timescales
means that the dynamics of the PR model is well approximated by two separate
models: one on the `slow' timescale that is appropriate when considering
changes on the timescale of the circadian and homeostatic processes such
as the timings of sleep and wake; and the other, the 
`fast' timescale,  which is appropriate when considering changes on
the timescale of the neuronal potentials such as the response to a night time 
disturbance. 
If the firing switching function $Q_j$ given in equation 
(\ref{eq:softswitch}) in the PR model is replaced by 
a hard switch, 
\begin{eqnarray}
Q_j = \begin{cases} 0		&\text{for  } V_j < \theta_S \\ 
\Qs		&\text{for  } V_j \geq \theta_S,
\end{cases}
\label{eq:hardswitch}
\end{eqnarray}
where $\Qs$ is the mean maximum firing rate of the neuronal population
and $\theta_S$ is the
value at which the switch occurs, we show in Appendix~\ref{app:BAPRswitch} that
the parameters for the slow dynamics of the PR model with a switch can be exactly mapped to
parameter values in the \BDB model, specifically,
\begin{eqnarray}
H_0^+  =  \frac{\theta_S + A_v + \nu_{vm} \Qs}{\nu_{vh}}, & & 
H_0^-  =  \frac{\theta_S  + A_v}{\nu_{vh}},
\nonumber \\
a  =  \frac{\nu_{vc}}{\nu_{vh}}, \qquad
\mu  =  \bar \mu \Qs, & & 
\chi_s  =  \chi_w = \chi. 
\label{eq:PRtoBA}
\end{eqnarray}
The lower threshold is therefore dependent on the mean drive to the VLPO 
and the threshold firing rate.
The difference between the thresholds in the \BDB model, 
$$
H_0^+ - H_0^-  =  \frac{\nu_{vm} \Qs}{\nu_{vh}}, 
$$
can then be interpreted physiologically as the amount by which the MA 
inhibits the firing of the VLPO during wake.  This makes intuitive sense:
there is hysteresis in the switch between wake and 
sleep because of the mutual inhibition between the MA and the VLPO.  In
the wake state, the VLPO requires a large drive to fire 
to counteract the inhibitory effects of the MA.  Once in the sleep state, less
drive is needed to maintain firing because the MA
is quiescent. 

\begin{figure}
\begin{center}
\includegraphics[scale=0.7]{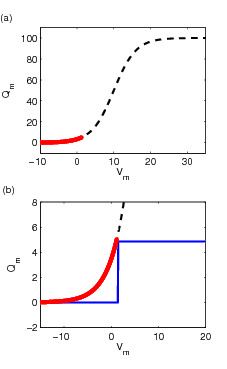}
\hspace*{1cm}
\includegraphics[scale=0.7]{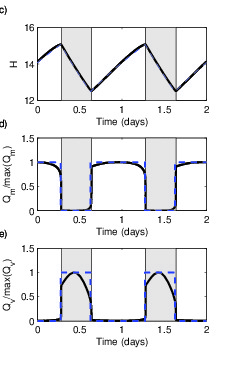}
\end{center}
\caption{(a) The dashed (black) line shows the firing function 
given by equation (\ref{eq:softswitch}); the thicker (red) line shows the portion 
that is used for the `normal' PR cycle. (b) A magnified version
of (a). The thin (blue) line shows the switch function (\ref{eq:hardswitch}). 
Panels (c),(d) and (e) show the
behaviour of the homeostat, $H$, and the firing rates $Q_m$ and $Q_v$
for the PR model (solid line) and the PR model with the hard switch 
(dashed line).  The switch parameters are $\theta_S=1.45$mV,
$\Qs=4.85$s$^{-1}$, the mean firing rate of the
  neural population during wake; all
other parameters are listed in Appendix~\ref{app:parameters}.
}
\label{fig:firingfunction}
\end{figure}

Using the standard parameters for the PR model, only a small part of the
firing function (\ref{eq:softswitch}) is used.  
This is illustrated in Figure~\ref{fig:firingfunction}(a),
where the firing function is shown by the dashed line and the typical range of 
values for $Q_m$ is 
shown by the thick line.  
We show in Appendix~\ref{app:PRswitchPR} that there is
a systematic way to relate the parameters for the original PR model 
to equivalent parameters
for the \BDB model that retain the timings and values at the extrema 
of the homeostat. 
In keeping with the fact that the mean firing rate
across the neuronal population $\Qs$ 
is much less than the maximum possible firing rate $\Qmax$, the 
value for $\Qs$ is significantly less
than $\Qmax$ but close to the mean firing rate across the
  population in the PR model: in fact
the actual firing function needed in the PR-switch model is shown by the thin line in 
Figure~\ref{fig:firingfunction}(b).  

\begin{figure}
\begin{center}
\includegraphics[scale=0.7]{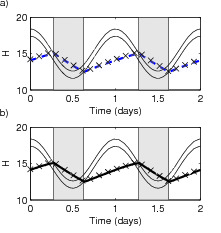}
\end{center}
\caption{(a) Comparison of the PR switch model
with the \BDB model. (b) Comparison of the PR model with the \BDB model.
Crosses show the \BDB model; solid line the PR model and (blue) dashed line
 the PR switch model.}
\label{fig:BAvsPR}
\end{figure}

Typical graphs of $H$ and $Q_j$ for both the original PR model and the
PR switch model are shown in Figure~\ref{fig:firingfunction}(c)-(e)
demonstrating the close agreement between the two cases.
Graphs comparing timeseries computed from the \BDB model 
and numerical integrations of the corresponding PR/PR switch model are 
shown in Figure~\ref{fig:BAvsPR}. 
The extremely good agreement of the two models is a result of the very
large disparity in timescales between the fast and slow systems. Consequently,
solutions of the PR model converge to solutions on the slow manifold on the 
timescale of minutes. 
Once on the slow manifold, solving the PR model is
essentially equivalent to solving the \BDB model.

In \cite{PR_JTB_08} it was recognised that the PR model could be
plotted in a similar way to the \BDB model, but the explicit
connection between parameters was not made.  It is stated that a key
difference is that in the \BDB model the value of $H$ remains between
the thresholds at all times, as in Figure~\ref{fig:BA_Daanparam}.
However, we note that this could be regarded as a matter of parameter
choice rather than a fundamental difference between the two models:
whether the \BDB model remains between the thresholds depends on the
relative gradients of the circadian and homeostatic processes at each
wake/sleep or sleep/wake transition.  
 Figure~\ref{fig:BAvsPR} shows that, with the PR parameters
used to model sleep regulation in humans, the two thresholds in
  the \BDB model are very close, hence the
circadian oscillation is the dominant sleep regulator and the two
  thresholds merge almost into one. 


\newpage
\section{The \BDB model re-visited}\label{sect:1dmap}

The link between the PR model and the \BDB model not only gives us a 
physiological interpretation of the thresholds in the \BDB model, it also
allows us to gain a greater insight into the dynamics of the PR model,
enabling understanding developed
in the context of the \BDB model to be interpreted in the physiological setting 
of the PR model.  
In this section, two different examples are discussed.  

\subsection{Transitions from monophasic to polyphasic sleep}

It is well-known that the \BDB model can show a range of different
sleep-wake cycles, including cycles that have multiple sleep episodes
each day, see Figure \ref{fig:BA_Daanparam}(a), and cycles that have a
period greater than one day, see Figure \ref{fig:BA_Daanparam}(c).
Indeed in \cite{Daan_84}, the authors postulate that
the \BDB model can explain the polyphasic sleep of many animals.  In
\cite{PRKA_PLoSCB_10}, it is shown that the sleep-wake cycles of many
different mammals can be understood by varying two parameters in the
PR model: the homeostatic time constant $\chi$ and the
constant component to the VLPO drive, $A_v$.  In the previous
  sections, we have demonstrated how the parameters of the PR model
  relate to those of the \BDB model, specifically, the homeostatic
  time constant~$\chi$ is present in both models and varying the drive
  to the VLPO~$A_v$ corresponds to varying the upper and lower
  thresholds without changing the distance between
  them. In~\cite{nakao_etal_im_1997,nakao_yamamoto_pcn_1998} it is
  shown that the \BDB
model can be understood as a one-dimensional map
  with discontinuities. In this section, we use this map to
     show how the observations
  in~\cite{PRKA_PLoSCB_10} and the postulate in \cite{Daan_84}
  are linked and clarify how the transition between
  different numbers of daily sleep episodes occurs.

\begin{figure}
\begin{center}
\includegraphics[scale=0.6]{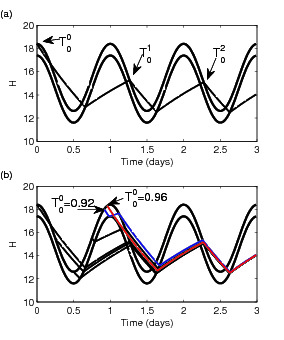}
\includegraphics[scale=0.6]{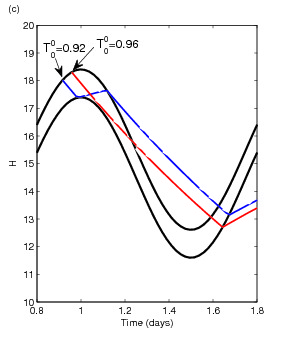}
\includegraphics[scale=0.6]{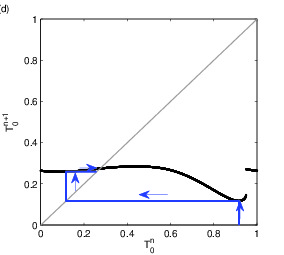}
\end{center}
\caption{(a) A single trajectory of the \BDB model showing successive
  times of sleep onset.  (b) Trajectories of the \BDB model for
  different initial sleep onset times. Each different sleep onset time
  results in a different sequence, $T_0^0, T_0^1\ldots$, but each
  sequence rapidly converges to the same sleep onset time, of
  $0.27$ modulo~1 day.  (c) A zoom of (b), showing only the
  trajectories for $T_0^0=0.92$ days and  $T_0^0=0.96$ days. 
  (d) First return map for the \BDB
  model.  
  The black line shows the return map, in other words
  for any value of sleep onset time on day $n$, $T_0^n$
  it shows the onset time of sleep on day $n+1$, $T_0^{n+1}$.
  The grey diagonal line is the line along which $T_0^n=T_0^{n+1}$.
  One typical trajectory is plotted for $T_0^0=0.92$ showing
  the rapid convergence to the periodic cycle where
  $T_0^n=T_0^{n+1}=0.27$ modulo~1 day, the point at which
  the return map and the diagonal line intersect.
  The discontinuity is a result of the fact that neighbouring values
  of $T_0^n$ exist that lead to very different values for $T_0^{n+1}$, as
  shown in (c).
  Parameter values
  for the \BDB model are based on the PR model for the human
    sleep-wake cycle and can be found in Appendix~\ref{app:PRswitchPR}.}
\label{fig:BAasmap}
\end{figure}

  First we introduce the one-dimensional map. Consider the \BDB
  model and suppose we start on the upper threshold, at time $T_0^0$,
  where the model switches from wake to sleep.  The dynamics of the
  \BDB model takes this starting point and, propagating it forward
  through one sleep and one wake episode, results in the next wake to
  sleep time, $T_0^1$, and then through a further sleep-wake episode
  to $T_0^2$ and so on, generating a sequence of sleep onset times
  $T_0^1, T_0^2, T_0^3 \ldots$.  This is illustrated for $T_0^0=0$ days in
  Figure~\ref{fig:BAasmap}(a).
  Different starting values $T_0^0$ generate different sequences of
  sleep times, as illustrated in Figure~\ref{fig:BAasmap}(b). For the
  parameter values chosen here, all sequences converge rapidly to the
  same monophasic periodic cycle.  A graphical way of understanding
  this sequence is to plot $T_0^{n+1}$ modulo 1 day against
  $T_0^n$ modulo 1 day (the first return map).  For any
  particular starting value, the sequence of iterates can then be
  found by drawing the cobweb diagram, as shown in
  Figure~\ref{fig:BAasmap}(d).  A monophasic sleep pattern corresponds
  to $T_0^{n+1}=T_0^n$ modulo 1 day and so corresponds to the
  intersection of the diagonal line with the map.  The fact that the
  sequences converge rapidly is related to the fact that the gradient
  of the map is close to zero for most values of $T_0^n$.  This rapid
  convergence means that a temporary change to timing of sleep will
  revert to the regular sleep-wake cycle within a few days.

  Phrasing the \BDB model in these terms
  illustrates that it can be represented as a
  one-dimensional map.  Probably the most well-known example of such
  maps is the logistic map \cite{May_76} which has been widely used to
  show that simple rules can lead to very complex dynamics.  A
  distinctive feature of the \BDB model is the fact that the map
  contains a discontinuity.  For the parameter values shown in
  Figure~\ref{fig:BAasmap}(d) this discontinuity occurs at
  $T_0^0\approx0.95$ days.  The discontinuity is a consequence of the fact
  that there exist neighbouring starting values $T_0^0$ that lead to
  trajectories that follow very different paths.  These occur whenever
  there are points that result in trajectories that become tangent to
  the thresholds.  For example, starting at $T_0^0=0.96$ days, the
  first sleep just misses the wake threshold at $1.08$ days so remains
  asleep until $1.6$ days resulting in a sequence $0.96, 1.6, \ldots$,
  as shown in Figures~\ref{fig:BAasmap}(b) and (c); whereas starting
  at the nearby value of $T_0^0=0.92$ days, the trajectory hits, rather
  than misses, the sleep threshold and the resulting sequence is
  $0.92, 1.1, \ldots$.

\begin{figure}
\begin{center}
\includegraphics[scale=0.55]{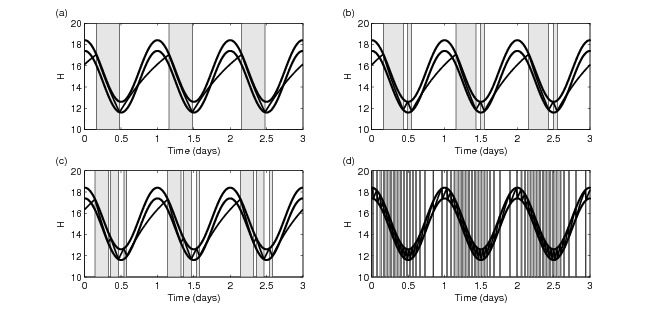}

\includegraphics[scale=0.55]{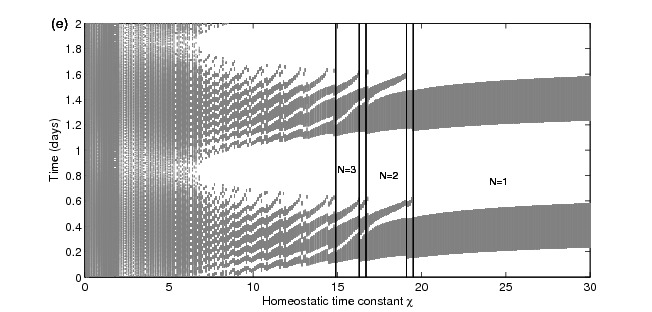}
\end{center}
\caption{Using the \BDB model with parameters as indicated in
    Appendix~\ref{app:PRswitchPR}, Figures (a)-(d) give 
    sleep-wake cycles for different values of 
the homeostatic time constant $\chi$ ($\chi=20,18,16,5$ hrs), illustrating
that reducing $\chi$ results in more daily sleep episodes.
(e) Sleep regions (shaded) as a function of $\chi$. Note that the
circadian maximum occurs at $t=0,1,\ldots$ days.
}
\label{fig:BAiteratedmappings}
\end{figure}

For the value of the clearance parameter $\chi=45$ hrs
used in Figure~\ref{fig:BAasmap},
the discontinuity does not have a significant impact on the
dynamics and
all trajectories converge rapidly to the same periodic cycle.  However, the
presence of the discontinuity is key to understanding the transition from
monophasic to polyphasic sleep. 
This is illustrated in Figure~\ref{fig:BAiteratedmappings}(a)-(d),
where a sequence of converged solutions to the \BDB model are shown for decreasing $\chi$. 
For $\chi=20$ hrs, the sleep-wake cycle is monophasic, but in the wake 
episode the trajectory comes close to, but just does not touch,
the upper threshold (Figure~\ref{fig:BAiteratedmappings}(a)).  If distance
from the upper threshold is a measure of sleepiness during wake, this would 
correspond to
a dip in alertness. If $\chi$ is reduced further, say to $\chi=18$ hrs as shown in
Figure~\ref{fig:BAiteratedmappings}(b), then the wake trajectory does not 
only come close to, 
it touches the upper threshold resulting in a short nap and a 
sleep-wake cycle that is bi-phasic with one longer sleep and one short sleep.  
Decreasing
$\chi$ further results in a sequence of further tangencies each of which
adds one additional sleep-wake episode.  Such transitions are known as
grazing bifurcations, tangent bifurcations,
or border collision bifurcations
and are characteristic of one-dimensional maps with discontinuities 
\cite{Nusse_Yorke_1992, LoFaro_1996,Banerjee_03}.  
In the return map, a grazing bifurcation occurs when the discontinuity in the 
map coincides with the diagonal line.  
They are responsible for period-adding transitions in the context of electronic circuits
and here, we see, are
responsible for sleep-episode-adding transitions.  
Such transitions have also been observed and analysed using
one-dimensional maps in the context of understanding the dynamics
of neurons \cite{LoFaroKopell_1999,CoombesOsbaldestin_2000}.

\begin{figure}
\begin{center}
\includegraphics[scale=0.6]{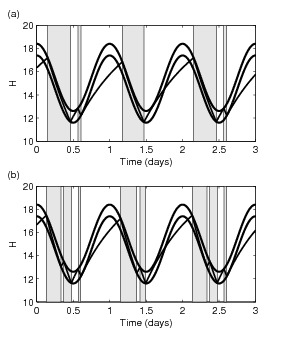}
\end{center}
\caption{Solutions of the \BDB model showing periodicity on the period
  of two days. (a) $\chi=19.3$ hrs (b) $\chi=16.6$ hrs. All other parameters
  are as in Figure~\ref{fig:BAiteratedmappings} and can be found in
  Appendix~\ref{app:PRswitchPR}.
}
\label{fig:BAintermediatetransitions}
\end{figure}

The sleep-wake pattern for varying $\chi$ is shown in
Figure~\ref{fig:BAiteratedmappings}(e).  For larger values of $\chi$
there is one episode of sleep each day: the model falls asleep exactly
once and always at roughly the same time ($N=1$).  A grazing
bifurcation occurs at around $\chi=19.8$ hrs and results in a region
between $15<\chi<19$ hrs where sleep is bi-phasic with one longer and one
shorter sleep each day ($N=2$).  A succession of further grazing
bifurcations take place as $\chi$ is reduced, resulting in increasing
numbers of daily sleep episodes.
From Figure~\ref{fig:BAiteratedmappings}(e) we see there are
intermediate regions between each value of $N$.  For example, between
the monophasic and biphasic region there is a small region around
$\chi=19.8$ hrs where the sleep pattern has a period of two days.  This
corresponds to a region where a grazing bifurcation has taken place,
causing an extra sleep period on one day, but this extra sleep period
is enough to mean that no additional sleep is needed on the following
day. The sleep wake trajectory in this case is shown in
Figure~\ref{fig:BAintermediatetransitions}(a).  Similar behaviour is
seen at each transition between different numbers of daily sleep
episodes and is characteristic of such transitions in one-dimensional
discontinuous maps \cite{Banerjee_06}: this is illustrated for the
transition between two and three sleep episodes in
Figures~\ref{fig:BAintermediatetransitions}(b) and 
    a similar pattern of sleep to that shown in Figure
\ref{fig:BA_Daanparam}(a) using parameters as in \cite{Daan_84}.
In fact, as shown for one-dimensional discontinuous maps in
\cite{Banerjee_06}, the situation is even more complicated:
in Figure~1 of [27] the first few layers of an infinite adding scheme
are set out.  This shows that, for example, the sequence of transitions from
sleeping once a day to sleeping twice a day is $\{1,1,1,\ldots\} \ldots
\{1,1,2,1,1,2\ldots\},$ $\{1,2,1,2\ldots\},$ $\{1,2,2,1,2,2,1\ldots\}$
$\ldots \{2,2,2,\ldots\}$, etc.
 
Further discussion of the map is given in the Supplementary Material.

\begin{figure}
\begin{center}
\includegraphics[scale=0.6]{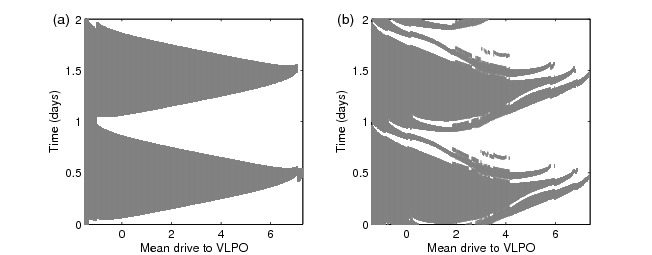}
\end{center}
\caption{Sleep timing in the \BDB model while simultaneously varying the upper and
    lower thresholds via $H_0^-= 1.45+A_v$ and $H_0^-= 2.46+A_v$
  with (a) $\chi=45$ hrs (b) $\chi=15$ hrs and
all other parameters as in Appendix~\ref{app:PRswitchPR}.  
Note that mean VLPO drive equals $c_0=A_v/\nu_{vc}$ for
consistency with \cite{PRKA_PLoSCB_10}.  Sleep regions are shaded.}
\label{fig:iteratemap_varyAv}
\end{figure}

 In \cite{PRKA_PLoSCB_10}, the behaviour of
    the PR model is examined both as the time constant $\chi$ and the
    mean drive to the VLPO, $A_v$ are varied.  Our parameter
equivalences between the PR model and the \BDB model (\ref{eq:PRtoBA})
show that increasing $A_v$ is equivalent to
increasing the upper and lower thresholds without
changing the distance between them.  One can then deduce for the
  \BDB model that for low
$A_v$, the homeostat will never reach the lower threshold and no wake
will occur.  Similarly, for high $A_v$ no sleep will occur.  For large
values of $\chi$ ($\chi$ greater than approximately $20$ hrs), the amount
of daily sleep varies approximately linearly with the mean drive to
the VLPO as shown in Figure~\ref{fig:iteratemap_varyAv}(a) and
  observed in~\cite{PRKA_PLoSCB_10}.  As seen before, the sleep-wake cycle is
monophasic and is largely independent of $\chi$ in this range.  The
actual transition between monophasic sleep and no sleep (or no wake)
occurs through grazing bifurcations, where this time the grazing
bifurcations result in periodic cycles that have wake (sleep) episodes
of greater than 24 hours: examples of such cycles are evident in
Figure~\ref{fig:iteratemap_varyAv}(a) at the extremes of the values of
$A_v$ that are shown.  For smaller values of $\chi$, where polyphasic
sleep exists, varying $A_v$ shows that, as the no sleep (or no wake)
threshold are approached, grazing bifurcations result in ever
decreasing numbers of sleep (wake) episodes until no sleep (no wake)
occurs, see Figure~\ref{fig:iteratemap_varyAv}(b).

In \cite{PRKA_PLoSCB_10}, it was shown that the sleep of many
mammalian species could be understood in the context of the PR model
by varying just two, physiologically plausible, parameters: $\chi$
  and $c_0=A_v/\nu_{vc}$.  Their results show: a sequence of
transitions from monophasic to polyphasic sleep as the time constant
$\chi$ is reduced but where total sleep daily sleep remains
approximately constant; for fixed $\chi$ and varying mean drive to the
VLPO a sequence of transitions from a state with no wake 
to a state with no sleep.  By using the relationship
between the PR model and the \BDB model we see that reducing $\chi$
results in a sequence of transitions from monophasic to polyphasic
sleep through grazing bifurcations that successively add sleep
episodes; at the transition between $N$ episodes of sleep and $N+1$
episodes of sleep, there are regions where sleep alternates between
$N$ and $N+1$ daily episodes (examples of such trajectories for the
PR model are shown in the Supplementary
Material). The parameter equivalences identified
in Section~\ref{sect:BAPRequivalence} show that
changing the mean drive to the VLPO is equivalent to simultaneously
shifting the upper and lower thresholds of the \BDB model.
The relation between the PR model and the \BDB model
  shows how this inevitably leads to grazing bifurcations and
  ultimately cycles with either no sleep or no wake.

The quantitative agreement is close with \cite{PRKA_PLoSCB_10}, but not exact: this
is because we have chosen a fixed value for $\mu$, the upper asymptote, 
in the \BDB model, the value to match the PR model
for $\chi=45$ hrs.  Varying $\chi$ in the PR model results in a small change to the precise 
region of the switching function that is used, which in turn induces some change in the 
value of $\Qs$.  Since $\mu=\bar \mu \Qs$ this results in some dependence of
$\mu$ on $\chi$ in the equivalent \BDB model.  One consequence is
that the switch from monophasic sleep to biphasic sleep occurs
at around $\chi=19$ hrs for the \BDB model instead of 
$\chi=16$ hrs for the PR model. More details can be found in the
  Supplementary Material.

\subsection{Wake effort}
Sleep deprivation experiments involve keeping subjects awake for an extended
period of time during
which cognitive and behavioural tests are undertaken to measure
sleepiness and performance.  One measure 
of sleepiness is 
the Karolinska Sleepiness Scale (KSS) score and in \cite{FPR_JTB_10}, 
the concept of `wake effort' is introduced for the PR model and good agreement between
wake effort and experimental
data on KSS scores is found.  Wake effort corresponds to a change in the drive to 
the MA and 
is interpreted as a need to provide the MA with greater stimulation in order to maintain
wake.  Here, we show how this can be re-interpreted in the context of the \BDB model.  

\begin{figure}
\begin{center}
\includegraphics[scale=0.75]{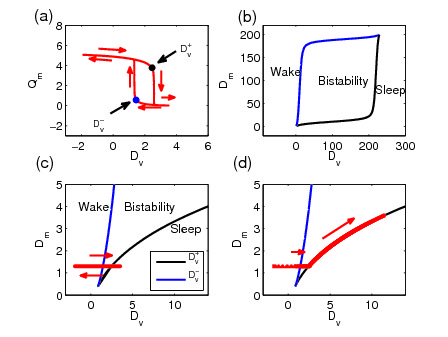}
\end{center}
\caption{(a) Sleep-wake cycle showing the MA firing rate $Q_m$ as a function 
of the drive to the VLPO $D_v$.  Over one cycle $D_v$ oscillates between high and
low values.  When
$D_v$ is low, $Q_m$ is high and the model is in a wake state.  When $D_v$ is
high, $Q_m$ is low and the model is in a sleep state.  The transitions from wake to
sleep and sleep to wake occur at $D_v^+$ and $D_v^-$ respectively.  
The size of the hysteresis loop depends on $D_m$, shrinking to nothing
for $D_m<0.4$ mV and for $D_m>200$ mV.
(b) The path of $D_m^+$ and $D_m^-$ in the $D_m,D_v$ plane. 
$D_m^+$ and $D_m^-$ do not exist for values of $D_m$ that are either less
than $0.4$ mV or greater
than $200$ mV. Consequently for $D_m<0.4$ mV or $D_M>200$ mV
increasing $D_v$ will result in a smooth change from high $Q_m$ (wake)
to low $Q_m$ (sleep) instead of the jump from one state to the other shown
in (a).
(c) A blow up of
(b), with the `normal' sleep-wake cycle superimposed. (d) The $D_m,D_v$ plane
showing the wake trajectory in a sleep deprivation experiment.}
\label{fig:saddlenodes}
\end{figure}

Wake effort in \cite{FPR_JTB_10} is presented by considering the graph of 
the MA firing rate $Q_m$ (or equivalently, $V_m$), against
the drive to the VLPO, $D_v$.  In a regular sleep-wake cycle, $Q_m$ follows a hysteretic
loop, see Figure~\ref{fig:saddlenodes}(a), where the transition from wake to 
sleep occurs close to $D_v^+$ and the transition from 
sleep to wake 
occurs close to $D_v^-$.  During sleep deprivation, 
it is argued in \cite{FPR_JTB_10} that by increasing $D_m$, rather than switch from wake to sleep, it is 
possible to stabilise the `ghost' of the wake state: the extent to which $D_m$ is increased
is known as the wake effort.  An alternative view of the
same idea is to consider the $(D_m, D_v)$-plane as shown in Figure~\ref{fig:saddlenodes}(b)
and recognise that $D_v^\pm$ are curves that divide
the parameter plane into regions where only the wake state exists, only the sleep
state exists, and a bistable region where both wake and sleep exist. There are also
regions for low $D_m$ ($D_m<0.4$ mV) and high $D_m$ ($D_m>200$ mV) where the two states cannot 
readily be distinguished.
The region of relevance for the parameters used in \cite{FPR_JTB_10} 
is close to the bottom of the bistable region, and is shown in 
blow-up in Figure~\ref{fig:saddlenodes}(c).  
The horizontal line represents the normal sleep-wake cycle: the time dependence
of the homeostatic and circadian processes result 
in $D_v$ oscillating backwards and forwards along the line, switching from
wake to sleep for increasing $D_v$ when $D_v=D_v^+$ and 
from sleep to wake for decreasing $D_v$ 
when $D_v=D_v^-$. 

In sleep deprivation experiments, subjects are prevented
from falling asleep at $D_v=D_v^+$.
At this point, in order to 
remain awake the only alternatives that keep the system in the wake region
are: decrease the drive to the VLPO, $D_v$; increase the drive to the MA, $D_m$ or
some combination of both of these.  In \cite{FPR_JTB_10},
it is argued that in order to maintain wake it is necessary to stimulate 
the MA, and therefore $D_m$ is increased to remain on the `ghost state', 
but this is equivalent to following the line $D_v^+$. 
The additional amount by which the MA is stimulated, the wake effort,
$W$ is then
$$
W  = D_m^+ - 1.3,
$$
where $D_m^+$ is a function of $D_v^+$ and is the solution of equations (\ref{eq:sn}) in 
Appendix~\ref{app:PRswitchPR}. 
For
the region of relevance shown in Figure~\ref{fig:saddlenodes}(c) and (d), the relationship
is close to linear with a small quadratic term and is well-approximated by
$$
D_m^+ \approx -0.012 D_v^{+^{2}} + 0.416 D_v^+ + 0.383.
$$ 

\begin{figure}
\begin{center}
\includegraphics[scale=0.7]{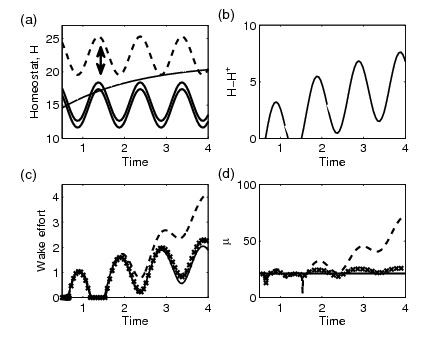}
\end{center}
\caption{(a) The \BDB model, showing the typical trajectory of the
  homeostatic pressure during a sleep deprivation experiment. Using
  the wake effort concept of \cite{FPR_JTB_10} suggests that the upper
  threshold moves simultaneously: the dashed line shows the position
  of the upper threshold after 4 days.  (b) The difference between the
  homeostatic pressure and the value at the `normal' threshold, $H-H^+
  = H(t) - (H_0^+ + aC(t))$.  (c) The wake effort computed from the
  \BDB model~\eqref{eq:BDBwake_effort} (solid line), the PR model as
  in \cite{FPR_JTB_10} using $\mu= \frac{\bar \mu Q_m^2}{\nu_h +
    Q_m^2}$ (crosses), the PR model with $\mu=\bar \mu Q_m$ (dashed
  line).(d) The dependence of  $\mu$, the upper asymptote, on
  time for the three different cases shown in (c). The downward
    spikes indicate that the model gets very close to falling asleep,
    hence $Q_m$ gets very close to~0.}
\label{fig:wakeeffort}
\end{figure}

In the \BDB model, acute sleep deprivation is modelled as a continued
increase in the homeostatic pressure.  In \cite{Daan_84} this is
interpreted as a suspension of the upper threshold, but with insight
gained from the the PR model, we see that an alternative
interpretation is that the upper threshold is continuously moved to
keep the model in the wake state, as shown in
Figure~\ref{fig:wakeeffort}(a).  The wake effort is then related to
the extent to which the threshold has to be moved, that is the
quantity $\max(H-H^+,0)$ with the upper threshold $H^+$ as given
  by~\eqref{eq:BAupperthreshold}. This quantity is shown in
Figure~\ref{fig:wakeeffort}(b).  Using the explicit relationships
between the parameters in the PR model and the \BDB model, the moving
of the threshold corresponds to a modified value for $D_v^+$ is given
by $D_v^+=H-H^++2.46$, as $\nu_{vh}=1$ and $2.46=\nu_{vh}H_0^+-A_v$,
  the value of $D_v^+$ if no wake effort is applied, so the wake
  effort for the \BDB model is
\begin{equation}\label{eq:BDBwake_effort}
W \approx 
-0.012 \left (H-H^+ \right )^2 + 0.357 \left ( H-H^+ \right ) .
\end{equation}
This resulting wake effort computed from the \BDB model
is shown by the solid line in Figure~\ref{fig:wakeeffort}(c) and
agrees very well with the calculation of the wake effort from the PR
model in~\cite{FPR_JTB_10} (crosses).

The close to linear relationship (the quadratic term has a very
  small coefficient) between wake effort in the PR model and $H-H^+$,
which is essentially the difference between the homeostatic pressure
and the circadian oscillator, demonstrates that the wake effort used
in \cite{FPR_JTB_10} is fundamentally similar to previous measures
used to compare performance and sleepiness scores.  The precise
scaling relationship and the degree of nonlinearity is dependent on
the shape of the bistable region in the $(D_m,D_v)$-plane shown in
Figure~\ref{fig:saddlenodes}(b), and on the choice of function for the
dependence of the homeostatic process on the firing rate of the MA.
In \cite{PR_JBR_07} and for many of the subsequent papers, the
  upper asymptote is give by $\mu=\bar
\mu Q_m$.  However, in \cite{FPR_JTB_10} the functional form $\mu =
\bar \mu \frac{Q_m^2}{\nu_h+Q_m^2}$ is used in order to "limit the
unrealistically high production rate at high $Q_m$''.  This change in
functional form has the effect of keeping $\mu$ approximately constant
during wake, which is why the agreement between the wake effort as
defined by \cite{FPR_JTB_10} agrees well with our analogous
computation from the \BDB model.  This is illustrated in
Figure~\ref{fig:wakeeffort}(c) and (d) where the wake effort and the
dependence of $\mu$ on time are shown for the \BDB model and for the
PR model with the two different functional forms for $\mu$.

The shape of the bistable region in the $(D_m,D_v)$ plane shows that
for $D_m$ larger than about
$30$ mV, there is a transition from relatively small changes in $D_m$ 
needed to maintain wake to very large changes in $D_m$ needed to maintain
wake; eventually it becomes impossible to maintain wake at all.
While for typical parameters used in the PR model this transition
occurs for infeasibly large values of $D_v$ and $D_m$, we note
that the shape of the bistable region is 
dependent on the parameters within the firing function and the 
choice of firing function itself.  Once fixed in \cite{PR_JTB_08}
these parameters have largely been left unchanged: we will
return to this point in the discussion. 

\section{Discussion}

The strengths of the \BDB model have been its inclusion of the two fundamental
processes that are believed to regulate the sleep-wake cycle along
with its graphical simplicity.  This has meant that it has been used extensively as a 
tool to understand the behaviour of the sleep-wake cycle, design experiments
and interpret data \cite{DijkandCzeisler_1995,Achermann_03}. 
A
weakness is the difficulty in relating the threshold levels of the homeostatic
pressure $H$ that
result in switches between wake and sleep to physiological quantities.

The PR model was developed with the same two governing processes in mind, but introduced
some physiological basis for the switching that occurs between wake and sleep.
In recent years, this model has been extensively tested in a range of scenarios, some of
which depend on the fast dynamics within the model, like the role of disturbances 
during sleep \cite{FPR_PRE_08}, but in many cases relying on the slow dynamics of the model.  
The four orders of magnitude between the neuronal timescale and the homeostatic/circadian
times scales means that the  timescale separation between the slow and fast dynamics is
very good.

Here we have shown that the slow dynamics of the PR model can be
explicitly related to the \BDB model, which provides new
  perspectives on both the \BDB model and the PR model.
Using this relationship, 
new insight into the meaning of the \BDB model has been gained.
Specifically, the distance between the thresholds is related to the
degree to which the MA inhibits the VLPO during wake and the values
of the thresholds are related to the parameters associated with the
modelling of the firing rates $Q_j$, the mean VLPO drive, and the
strength of the homeostat.  The parameter comparison also highlights
the fact that there is no strong reason why the homeostatic pressure
should remain between the thresholds in the \BDB model, see for
  example Figure~\ref{fig:BAvsPR}.  For values 
between the thresholds, either sleep or wake can occur.  Above the
upper threshold, only sleep can occur: this could be viewed as a
region of  sleep, from which it is hard to
wake up.
Below the lower threshold, only wake can occur, representing
times when it is difficult to fall asleep.


Motivated by the strong relationship between the \BDB
    model and the slow dynamics of the PR model, we
have re-visited the \BDB model.  By using that
the \BDB model can be represented as a one-dimensional map with
discontinuities we are able to interpret the transitions from
monophasic to polyphasic sleep as grazing bifurcations.  This provides
the dynamical underpinning for the observation that the PR model gives
a systematic framework which encompasses many different mammalian
species and confirms the hypothesis of \cite{Daan_84} that such a
framework could be present in the \BDB model.  Furthermore, it
suggests that `typical' transitions with varying clearance parameter,
at least for the larger mammalian species with relatively large
clearance parameters, will involve gaining or losing one sleep episode
a day.  We note that the sequence of transitions for increasing $\chi$
is consistent with observations of changes in the daily sleep patterns
of early childhood.

Varying the homeostatic time constant as shown in
Figure~\ref{fig:BAiteratedmappings}(a) suggests that for large mammals
(large $\chi$)  sleep
regulation is dominated by the circadian rhythm.  In contrast, as
shown in Figure~\ref{fig:BAiteratedmappings}(d), small mammals are more
strongly driven by their metabolism and it is the homeostatic
component that dominates. 
However, we note that the
equivalence of the two models raises some interesting questions on accepted parameter values: in
both models the homeostatic process is modelled in a similar way, with
exponential decay during sleep and an exponential approach to an upper
asymptote during wake.  
In the context of the two process model, accepted physiological markers for
the homeostatic process are slow waves in the sleep EEG and
theta activity in the EEG during wake respectively, both of which
are readily measured.  The
time constants $\chi_s$ and $\chi_w$ differ during wake and sleep and are measured to
be $\chi_s\approx 4 $ hrs and
$\chi_w\approx 18 $ hrs in humans \cite{BorbelyAchermann_99}.
An important physiological question is the necessity for two
different time constants for the homeostatic process,  one
for wake and one for sleep. Animal  \cite{Frankenetal_2001} and human
experiments \cite{Jennietal_2005} strongly suggest that the time constant during
wakefulness varies with genetic background (animals)  and during
development (humans)  whereas the time constant during sleep appears
more invariant within species.
In the context of
the PR model, the homeostatic process represents the concentration of
somnogenic factors such as adenosine, which are not easily accessible.
During wake, adenosine is produced
more quickly in the brain than it is cleared, decreasing the inhibition to the VLPO.
A single value $\chi=\chi_s=\chi_w=45 $ hrs is taken in order to replicate
typical sleep patterns for adult humans.   Given that in both models,
the homeostatic process plays a key role in determining patterns of sleep
and wake,
it would be interesting to extend the modelling of the homeostatic process in the
PR model to allow $\chi_s$ and $\chi_w$ to differ and determine whether a
different parameterization of the PR model would lead to time constants
in-line with measured values for the two process model.


The grazing bifurcations have been shown to occur as the clearance
parameter $\chi$ and as the mean drive to the VLPO or
  equivalently, both the upper and lower thresholds, are 
  simultaneously varied.
However, it is clear that the tangencies between the sleep-wake
trajectories and the thresholds that give rise to these bifurcations
could also occur if the distance between the
thresholds (see \cite{nakao_etal_im_1997,nakao_yamamoto_pcn_1998})
or the upper and lower asymptotes of the homeostatic process are
varied.  A systematic study will be carried out elsewhere.

The \BDB model has been compared with sleep deprivation experiments by
assuming that the upper threshold is no longer present and that the
sleep pressure continues to increase, with sleepiness linearly related
to the difference between the homeostat and the circadian process.
Here, we have demonstrated that the notion of `wake effort' introduced
in \cite{FPR_JTB_10} is a similar measure and is equivalent to
imagining, not that the upper threshold has vanished, but that
increasing the stimulation to the MA results in increasing the upper
threshold in line with the increase in $H$.

Similarly, one could also imagine a `sleep effort' that would be required to keep the
model asleep when it would naturally wake.  This could be achieved by reducing the lower
threshold in the \BDB model or, equivalently, decreasing the stimulation to the MA, $D_m$.  
As can be seen from Figure~\ref{fig:saddlenodes}(b), the PR model parameters 
suggest that, while it is possible to extend the wake state significantly by increasing
$D_m$, the capacity to extend the sleep state is more restricted.  This observation is
sensitive to the precise parameters and definition of the firing function.
The asymmetry between sleep and wake 
is equivalent to the fact that in~\cite{FPR_JTB_10}, 
the authors noted that the `sleep ghost' is
less prominent than the `wake ghost'.


The equivalence between the PR model on the slow timescale and the
\BDB model is exact when the firing function is a hard switch, but
when the firing function is sigmoidal is more subtle.  This is
because, in the PR model, the upper/lower asymptotes of the
homeostatic process are modelled as a  a
  function of~$Q_m$ the firing rate of the MA.
    With a hard switch, $Q_m$ takes only two values, $\Qs$
or zero (similar to the \BDB model), but with a sigmoid it varies
continuously.  Except in the neighbourhood of bifurcations, for
monophasic sleep we have shown that one can fix the maximal value 
of $Q_m$ and the switching
  voltage~$\theta_S$ such that the times when the homeostatic
pressure reaches its extreme values in the PR and \BDB models
co-incide. The precise values of $Q_m$ and $\theta_S$ needed, and
therefore the values of the asymptotes in the equivalent \BDB model,
depend to some extent on the other parameters in the model.  In this
paper we have taken the approach of fixing the values of the
asymptotes as those needed to agree with the PR model for their
`normal' values of the parameters at $\chi=45$.  We have not then
varied the asymptotes as other parameters are changed which means that
the quantitative agreement between the results from the \BDB model and
the PR model are not exact.  Nevertheless, the sequence of transitions
and the underlying mechanism through grazing bifurcations carry over
between the two models with only minor quantitative differences.  In
the case of the wake effort, the dependence of the upper
asymptote on the firing rate in the PR model
means that there is approximately a 10\% difference in the wake effort
between the \BDB and PR models after four days.

However, the fact that implicit in the PR model is a non-constant
asymptotic value for the homeostatic process has wider implications.
Sleep deprivation experiments tend to show a leveling off of
psychomotor vigilance test (PVT) scores over a period of a few days,
similar to the levelling off seen in the wake effort shown in Figure
\ref{fig:wakeeffort}.  In contrast, chronic sleep restriction
experiments, where subjects repeatedly are allowed less sleep than
they need, tend to show a linear increase in PVT over the timescale of
typical experiments.  In order to explain this,
in~\cite{Avinash_05}, Avinash {\em et al}  considered a \BDB model
but suggested that the upper and lower asymptotes varied with time.
This idea was generalised in \cite{McCauley_09}.  Both papers suggest
that the time variation occurs through some longer timescale process.
We note that within the context of the PR model, during sleep
deprivation or chronic sleep restriction the values of the firing
function will tend to increase, automatically inducing some time
dependence in the values of the asymptotes.

The asymptotes and therefore the wake effort in the PR model are
sensitive to the particular choice of the firing function and the
functional dependence of the upper asymptote on $Q_m$.  For parameter
choices made in \cite{FPR_JTB_10}, $Q_m$, like $D_m$, depends
approximately linearly on wake effort.  However, note that the shape
of the relation between $D_v$ and $D_m$ shown in
Figure~\ref{fig:saddlenodes} means that for high $D_m$ there is a
`corner' where to stay awake longer means that a very large increase
in $D_m$ is needed.  This transition suggests that a critical change
in behaviour for large wake effort, although it is unclear whether
this could give an alternative explanation for the behaviour at
extreme sleep restriction to the `bifurcation' suggested by
\cite{McCauley_09}.  This corner can be further understood by
re-examining the firing function shown in
Figure~\ref{fig:firingfunction}.  Since only a small part of the
sigmoid is used under `normal' conditions for the PR model, increasing
$D_m$ will result in an almost linear change to the range of $Q_m$.
However, once $D_m$ is large, it becomes increasing difficult to
increase $Q_m$ by increasing $D_m$ and the corner in
Figure~\ref{fig:saddlenodes} corresponds to the flattening off of the
relationship between $Q_m$ and $D_m$.  While this is beyond the
physiological range of the parameters, this part of the PR model has
been less constrained by physiological parameters or behaviour than
many other features of the model and a slightly different firing
  function could lead to a corner at more physiological values.  The
relationship between the \BDB based model
in~\cite{McCauley_09}, the PR model and the modelling of sleep
deprivation versus sleep restriction deserves further attention and
will be the subject of a future paper.


In order to better understand sleep/wake regulation it is essential
that models that incorporate neurophysiology are developed, analysed and used.
However, as models become more complex two problems arise.  Firstly they
become difficult to analyse systematically, with large numbers of numerical
simulations becoming the principle method used to establish the behaviour of the system.
Secondly, there is a proliferation of parameters
which cannot be easily determined experimentally.  One consequence is
that it becomes difficult to establish the relative merits of different models.
By demonstrating that the \BDB model and the PR model are essentially
the same for sleep-wake phenomena on the slow time-scale of hours we
have not only gained insight on the interpretation of both models but also
established the mechanism for transitions between different patterns of
sleep and wake in the PR model.    This link also suggests some
interesting avenues for future extensions of the PR model based on
recent insights and research on the two-process and related models.

\section*{Acknowledgements}
This work was partially supported by the Engineering and Physical Sciences Research
Council (grant number EP/I000992/1) and by the Royal Society of Medicine (DJD).

\bibliographystyle{unsrt}
\bibliography{sleep}

\newpage

\appendix
\section{Parameter values} \label{app:parameters}

\mbox{}

\begin{table}[ht]
\centering \tabcolsep3.8pt 
\begin{tabular}{|ccc|}
\hline\hline
 Parameter & PR & PR switch  \\
 \hline\hline
$\Qmax$ or $\Qs$  &  $100$s$^{-1}$ & $ 4.85$s$^{-1}$ \\
 $\theta$   & $10$mV         & $1.45$mV      \\
 $\sigma$'  & $3$mV          &  -            \\
 $\nu_{vm}$ & $2.1$mVs       & $ 0.208$mVs      \\
 $\nu_{mv}$ & $1.8$mVs       & $1.8$mVs      \\
 $\nu_{vc}$ & $2.9$mV        & $2.9$mV      \\
 $\nu_{vh}$ & $1$mVnM$^{-1}$ & $1$mVnM$^{-1}$\\
 $A_m$      & $1.3$mV        & $1.5$mV       \\
 $A_v$      & $13.05$mV      & $13.05$mV     \\
 $\tau_m$   & $10$s          & $10$s        \\
 $\tau_v$   & $10$s          & $10$s        \\
 $\chi$     & $45$hrs        & $45$hrs      \\
 $\bar \mu$      & $4.4$nMs        & $4.4$nMs     \\
\hline
\end{tabular}
\caption{Typical parameter values for the PR
  model and the equivalent parameters for the PR model with a hard switch:
  these are needed to find appropriate parameter values for the \BDB model.
  Further details on how to find values of $\nu_{vm}$ and 
  $Q_S$ are given in the Supplementary  Material.
  All parameters have been defined to be positive, consequently some of the
  signs in equations (\ref{eq:PRmodel}) are opposite to their original
  definitions in~\cite{PR_JBR_07}. The mean component of the
  circadian drive in the PR model   has been incorporated in the
  definition of $A_v$, $A_v=\nu_{vc} c_0$, $c_0=4.5$.} 
\label{tab:paramtable}
\end{table}

\section{PR switch to \BDB comparison}\label{app:BAPRswitch}

The equations for the PR switch model are 
\begin{eqnarray}
\tau_v \dot{V}_v +V_v &=&  - \nu_{vm} \Qs {\cal H}(V_m-\hat \theta_S)+D_v(t) \nonumber \\
\tau_m \dot{V}_m +V_m &=&  - \nu_{mv} \Qs {\cal H}(V_v-\hat \theta_S)+D_m(t) \nonumber \\
\chi \dot{H}  +H & = & \bar \mu \Qs {\cal H}(V_m-\theta_S),
\label{eq:PRswitch}
\end{eqnarray}
where
\begin{eqnarray*}
D_v &=& \nu_{vh} H - \nu_{vc} C(t) -A_v\\
D_m &=& A_m. 
\end{eqnarray*}

Since $\tau \ll \chi$ we introduce the small
  parameter~$\epsilon={\tau}/{\chi}$, the fast time $\hat t=t / \epsilon$
and the slow time $T=t$, $d/d\hat t =
\epsilon\,d/d t$ and  $d/dT=d/dt$.
Then, at $O(1)$ (slow time) equations (\ref{eq:PRswitch}) become
\begin{eqnarray}
 V_v & = & -\nu_{vm} \Qs {\cal H}(V_m-\theta_S)+D_v(T) \nonumber \\
 V_m &=&  -\nu_{mv} \Qs {\cal H}(V_v-\theta_S)+D_m(T) \nonumber \\
\chi \dot{H}  +H & = &  \bar \mu \Qs {\cal H}(V_m-\theta_S),
\end{eqnarray}
where
\begin{eqnarray*}
D_v &=& \nu_{vh} H - \nu_{vc} C(T) - A_v\\
D_m &=& A_m.
\end{eqnarray*}
During wake, these have solution
\begin{eqnarray*}
 V_v & = & - \nu_{vm} \Qs + \nu_{vh} H - \nu_{vc} C(T) - A_v,\\
 V_m &=&   A_m, \\
H & = &  \bar \mu \Qs + \left( H_0 - \bar \mu \Qs \right ) e^{(T_0-T)/\chi}.
\end{eqnarray*}
During sleep, these have solution
\begin{eqnarray*}
 V_v & = &   \nu_{vh} H  - \nu_{vc} C(T) - A_v,\\
 V_m &=&  -\nu_{mv} \Qs  + A_m, \\
 H & = &  H_0 e^{(T_0-T)/\chi}.
\end{eqnarray*}
Transitions between wake and sleep when $V_m=\theta_S$,
so the switch from wake to sleep occurs when
$$
H \equiv H^+= \frac{\theta_S + A_v + \nu_{vm} \Qs + \nu_{vc}C(T) }{\nu_{vh}},
$$
and from sleep to wake when
$$
H \equiv H^-= \frac{\theta_S  + A_v + \nu_{vc}C(T) }{\nu_{vh}}.
$$
By comparison with equations (\ref{eq:BAsleep})-(\ref{eq:BAlowerthreshold}) 
we see that the \BDB model and the dynamics of the
PR switch model on the slow manifold are 
equivalent if 
\begin{equation}\label{eq:PR_switch_BA_thresholds}
H_0^+  =  \frac{\theta_S + A_v + \nu_{vm} \Qs}{\nu_{vh}}, \qquad
H_0^-  =  \frac{\theta_S  + A_v}{\nu_{vh}}, 
\end{equation}
$$a  =  \frac{\nu_{vc}}{\nu_{vh}}, \qquad
\mu  =  \bar \mu \Qs, \qquad
\chi_s  =  \chi_w = \chi. 
$$
For the values of the PR parameters listed in Appendix A and
used in Figure~\ref{fig:BAvsPR}, 
$$
H_0^+  =  15.5, \quad 
H_0^-  =  14.5, \quad 
a  =  2.9, \quad
\mu  =  21.4, \quad
\chi_s  =  45 \mbox{hrs}, \quad
\chi_w  =  45 \mbox{hrs}.
$$
It is also necessary to take $A_m>\theta_S$, otherwise no switching occurs.

\section{PR to \BDB comparison}\label{app:PRswitchPR}
On the slow manifold, the PR model is
\begin{eqnarray*}
V_v & = & - \nu_{vm} Q_m + D_v(T) \\
V_m & = & - \nu_{mv} Q_v + D_m \\
\chi \frac{dH}{dt} + H  & = & \bar \mu Q_m \\
\end{eqnarray*}
where $Q_j, j=m,v$ is given by equation (\ref{eq:softswitch}).  
For a fixed value of $D_v$ these have one or three solutions,
with the transition between one and three solutions happening
at saddle-node bifurcations, $D_v^\pm$ that satisfy
\begin{eqnarray}
D_{v}^\pm & = & V_v-\nu_{vm} Q_m \nonumber \\
D_m^\pm & = & V_m-\nu_{mv} Q_v \label{eq:sn}\\
\frac{\nu_{vm} \nu_{mv}}{\sigma'^2} 
 & = & \left ( Q_v - \Qmax \right ) \left ( Q_m - \Qmax \right ). 
 \nonumber
\end{eqnarray}
The values of
$D_{v}^\pm$ depend on $\nu_{vm}, \nu_{mv}, \Qmax, \sigma'$ and $\theta$,
and for the values commonly used in the PR model and listed in Table~\ref{tab:paramtable}
give  $D_v^+ = 2.46$ and $D_v^- = 1.45.$   

The sleep-wake cycle corresponds to slowly changing $D_v$, tracing out a
path on the slow manifold as shown in Figure~\ref{fig:saddlenodes}(a).  Transitions
from wake to sleep and from sleep to wake occur close to $D_v^+$ and $D_v^-$
respectively.  
In order to find parameter values that retain the maximum and minimum
values and timings for the homeostatic process for monophasic sleep
away from bifurcation points
the following algorithm is followed:
\begin{itemize}
\item First the identification between the threshold values
and the saddle node bifurcations in the PR model is made, leading to
$$
H_0^+ = \frac{D_v^+ + A_v}{\nu_{vh}}, \quad
H_0^- = \frac{D_v^- + A_v}{\nu_{vh}},
$$
then
$$
H_0^+ - H_0^- = \frac{D_v^+-D_v^-}{\nu_{vh}}.
$$
\item Numerically integrating the PR model during monophasic sleep results in trajectories
for the homeostat that increase to a maximum during wake and decrease to a minimum
during sleep.  The maximum and minimum values occur close to the switches from wake
to sleep and sleep to wake respectively.
During wake, the
BDB model gives
$$
H(t) = \mu + \left (H_{\rm min} - \mu \right) {\rm e}^{(t_{\rm min}-t_{\rm max})/\chi_w}. 
$$
Hence, taking
$$
\mu = \frac{H_{\rm max} - H_{\rm min} \exp 
  \left ( \frac{t_{\rm min} - t_{\rm max}}{\chi_w} \right )}{1 - \exp \left 
  (\frac{t_{\rm min} - t_{\rm max}}{\chi_w}
 \right )}
 = 21.35,
$$
results in a trajectory for the BDB model that passes through the required values at the
required times.
\item One can do a similar matching for the decreasing $H$ phase to find a value for
the lower asymptote. For the simulations presented here, the value of zero was taken
for the lower asymptote.
\end{itemize}

By integrating the PR model with the `normal'
parameter values listed in Appendix A, it is found that the minimum occurs
at $H_{\rm min} = 12.51, t_{\rm min} = 15.31$ hours and the maximum
at $H_{\rm max} = 15.07, t_{\rm min} = 30.67$ hours.
\begin{itemize}
\item Comparing the expression for $H_0^-$ above with that for the PR switch
model in~\eqref{eq:PR_switch_BA_thresholds}, gives
$$
\theta_S = D_v^-.
$$
\item The relation for $\Qs$ in~\eqref{eq:PR_switch_BA_thresholds}
gives
\[
\Qs = \frac{\mu}{\overline\mu}.
\]
\item Considering $H_0^+-H_0^-$ as given above with $H_0^+$ and $H_0^-$ as
in~\eqref{eq:PR_switch_BA_thresholds}, leads
to
$$
\nu_{vm}\Qs = D_v^+-D_v^-.
$$
\end{itemize}
Hence for the typical values of the PR parameters listed in
Table~\ref{tab:paramtable} and used in Figure~\ref{fig:BAvsPR},
$$
\theta_S=1.45, \quad\Qs=4.85\, \quad \nu_{vm} = 0.208\,.
$$
It is also necessary to take $A_m>\theta_S$ in the PR switch model,
otherwise no switching occurs.
%
%


\end{document}


\title{Mathematical Models for Sleep-Wake Dynamics: Comparison of the Two-Process Model
and a Mutual Inhibition Neuronal Model\\
\mbox{}\\
\bf Supplementary material
}
\author{A.C.~Skeldon$^*$, D.-J.~Dijk$^\dag$, G.~Derks$^*$ \\
$^*$ \small Department of Mathematics, University of Surrey, Guildford, Surrey, GU2 7XH, \\
$^\dag$ \small Faculty of Health and Medical Sciences, University of Surrey, Guildford, Surrey, GU2 7XH.
}
\maketitle

\subsection*{The \BDB model as a one-dimensional map}

As discussed in Section 4, the \BDB model can be represented as
a one dimensional map with discontinuities.  
The map is given
by
\begin{equation}
T_0^{n+1} = G(T_0^n)
\end{equation}
where $G(T_0^n)$ is defined as follows. For $t\in(T_0^n,T_w^n]$
\begin{eqnarray*}
H(t) & = & \left ( H^+_0+a C(T_0^n) \right ) {\rm{e}}^{(T_0^n-t)/\chi_s},
\\
 & & \qquad T_w^n = \min(t) \quad \mbox{s.t.} \quad H(t)= H^-_0 + a C(t).
\end{eqnarray*}
For $t\in(T_w^n,T_0^{n+1}]$
\begin{eqnarray*}
H(t) & = & \mu + \left (
\left ( H^-_0 + a C(T_w^n) \right) - \mu \right) {\rm{e}}^{(T_w^n-t)/\chi_w},\\
&& \qquad T_0^{n+1} = \min(t) \quad \mbox{s.t.} \quad H(t)= H^+_0 + a C(t),
\end{eqnarray*}
where $ C(t)  =  \cos(\omega (t-\alpha))$; $H_0^\pm$ are the mean values for the upper/lower
thresholds; $\mu$ is the upper asymptote; $\chi_s$ and $\chi_w$ are the time constants for
the homeostatic process during sleep and wake; $a, \omega$ and $\alpha$ are the amplitude,
frequency, phase shift, of the circadian process $C(t)$.

In [27] a detailed analysis of piecewise-linear discontinuous maps
is presented.  It is argued that these represent the normal forms for many systems
in the neighbourhood of the discontinuity and that three different bifurcation
scenarios are observed in such systems.  Of the three scenarios, the 
particular case that is relevant
for the \BDB model parameters used to match the PR model in this paper 
is the scenario labelled as period adding
bifurcations.  It is not yet clear whether the two other scenarios, which 
[27] label as period increment with coexistence of attractors and
pure period increment scenarios, can also occur.   

The sequence of bifurcations
shown in Figure 6(e) is typical of period adding bifurcations.  A
more conventional way to present the bifurcation diagram is shown in 
Figure S1 where the values of $T_0^n$ are plotted against
the bifurcation parameter.  The term `period-adding' refers to the fact that 
the period of the iterated map changes as a function of the parameter. So,
for example, when the number of daily sleep episodes changes from one to two,
the map repeats itself after two iterations.   

In the main body of the text it is highlighted that between parameter values
where solutions with $N$ daily sleep episodes and solutions with $N+1$ daily
sleep episodes there are solutions which alternate between $N$ and $N+1$ sleep
episodes, as shown in Figure~7. The same
sequence occurs in the PR model, as illustrated in 
Figure~S2.

\begin{figure}
\begin{center}
\includegraphics[scale=0.55]{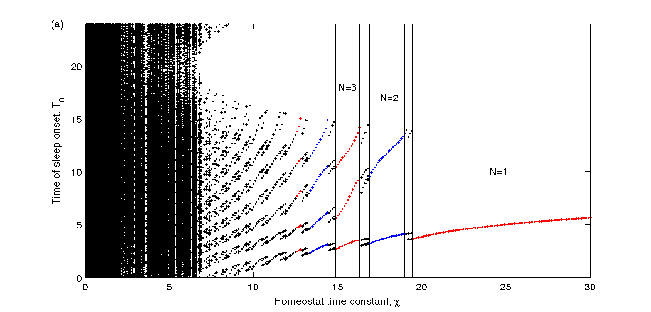}
\end{center}
\caption{ Bifurcation diagram showing the times of sleep onset against
  the clearance parameter $\chi$ in the \BDB model. Regions of
  parameter space where one, two and three daily sleep episodes ($N$)
  are delineated but not for higher values for clarity.  }
\label{fig:BAmap_supp}
\end{figure}

Yet another way of presenting the bifurcation diagram for the iterated
map is to plot the length of the daily sleep episodes.  This is shown
in Figure~S3.  On this diagram is also plotted the mean daily total
sleep.  This shows that for the \BDB model, as for the PR model, the
mean total daily sleep is approximately independent of the homeostatic
time constant.
\begin{figure}[h]
\begin{center}
\includegraphics[scale=0.6]{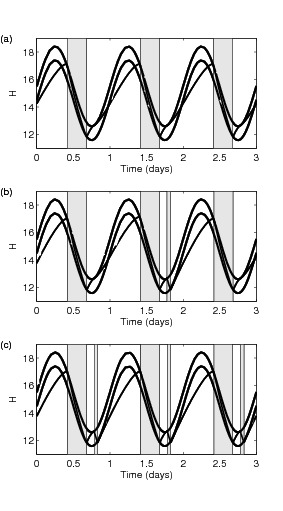}
\end{center}
\caption{PR model transitions from monophasic to biphasic sleep patterns as $\chi$ is 
reduced. (a) $\chi=16$ hrs,(b) $\chi=15.9$ hrs, (c) $\chi=15.8$ hrs.}
\label{fig:PRintermediatetransitions}
\end{figure}

\begin{figure}
\begin{center}
\includegraphics[scale=0.55]{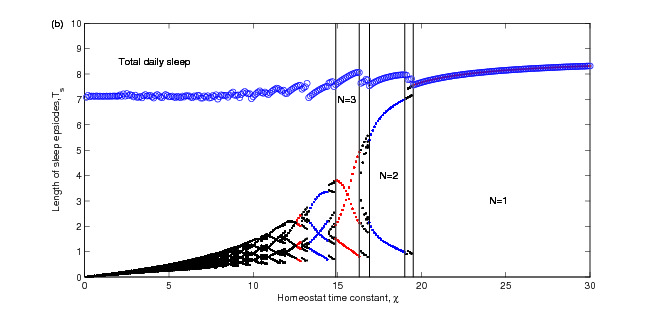}
\end{center}
\caption{
Bifurcation diagram showing length of sleep episode in the \BDB model as a
function of the parameter~$\chi$.  So, for example, at $\chi=22$ hrs there is one
single daily sleep episode of length approximately $8$ hrs and for
$\chi=18$ hrs there are two daily sleep episodes of approximately $6.6$ and
$1.5$ hrs respectively. 
}
\end{figure}

\subsection*{The upper asymptote}

\begin{figure}
\begin{center}
\includegraphics[scale=0.6]{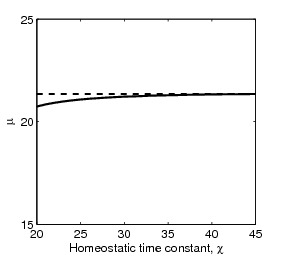}
\end{center}
\caption{\gd{Values of the upper asymptote $\mu$ in the \BDB model that are needed to fit the
PR model for monophasic sleep.}
}
\label{fig:mudependence}
\end{figure}

In Appendix C, it is shown how to fit the parameter $\mu$ of
the \BDB model such that the homeostatic switching happens with the
same timing as in the PR model. This fitting is dependent on the
sleep-wake cycle and if $\chi$ varies, this sleep wake cycle will
slightly vary and lead to different timings of the homeostat. In its
turn, this would lead to a correction for the paremeter $\mu$.   
In  Figure~\ref{fig:mudependence} the dependence of the
parameter $\mu$ is depicted for monophasic sleep. 
